\documentclass[a4paper]{jpconf}

\newcommand{\apj}{ApJ}
\newcommand{\apjl}{ApJ\xspace}

\newcommand{\aap}{A\&A}

\newcommand{\mnras}{MNRAS\xspace}

\newcommand{\ssr}{Space Sci. Rev.}

\newcommand{\solphys}{Sol. Phys.}

\newcommand{\grl}{Geophys. Res. Lett.}
\newcommand{\jgr}{J. Geophys. Res.}

\newcommand{\jswsc}{J. Space Weather Space Clim.}
\usepackage{xspace}
\usepackage{graphicx}
\usepackage{bm,url}
\usepackage{amsmath,txfonts}
\usepackage[dvipsnames]{xcolor}

\newcommand{\vect}[1]{\mathbf{#1}}
\renewcommand{\vec}{\vect}
\newcommand{\td}[2]{\frac{d #1}{d #2}}

\newcommand{\pd}[2]{\frac{\partial#1}{\partial#2}}

\newcommand{\pdd}[2]{\frac{\partial^2#1}{\partial#2^2}}
\usepackage{fancyhdr}
\begin{document}
\title{Spreading protons in the heliosphere: a note on cross-field diffusion effects}

   \author{N. Wijsen$^{1,2}$, A. Aran$^{2}$, J. Pomoell$^{3}$ and S. Poedts$^{1}$}
\address{$^{1}$ Department of Mathematics/Centre for mathematical Plasma Astrophysics, KU Leuven, Belgium}
\address{$^{2}$  Departament F\'{i}sica Qu\`antica i Astrof\'{i}sica, Institut de Ci\`encies del Cosmos (ICCUB), Universitat de Barcelona (IEEC-UB), Spain}
\address{$^{3}$  Department of Physics, University of Helsinki,  Finland}

\ead{nicolas.wijsen@kuleuven.be}

\begin{abstract}
We study how a high-speed solar wind stream embedded in a slow solar wind affects the transport and energy changes of solar energetic protons in interplanetary space, assuming different levels of cross-field diffusion.  
This is done using a particle transport model that computes directional particle intensities and first order parallel anisotropies in a background solar wind  generated by the magnetohydrodynamic model EUHFORIA.  
In particular, we consider a mono-energetic 4 MeV proton injection over an extended region located at a heliographic radial distance of 0.1 AU.  
By using different values for the perpendicular proton mean free path, we study how cross-field diffusion may affect the energetic particle spread and intensity profiles near a high-speed solar wind stream and a corotating interaction region (CIR). 
We find that both a strong cross-field diffusion and a solar wind rarefaction region are capable of dispersing SEPs efficiently, producing overall low particle intensities which can in some cases prevent the SEPs from being detected in-situ, since their intensity may drop below the detected pre-event intensity levels. 
We also discuss how accelerated particle populations form on the reverse and forward shock waves, separated by the stream interface inside the CIR.
 Under strong levels of cross-field diffusion, particles cross the SI and hence both accelerated particle populations merge together.

\end{abstract}

\section{Introduction}

Solar energetic particles (SEPs) are electrons, protons and heavy ions that have been accelerated to supra-thermal energies (up to a hundred of MeV or even $\sim$~GeVs, in the case of ions), typically during solar eruptive events. 
These charged particles may for example be energised at solar flare sites during magnetic reconnection events, or at shock waves driven by coronal mass ejections (CMEs).  
After escaping from their acceleration site, SEPs propagate through the inner heliosphere, guided by the interplanetary magnetic field (IMF).  
On their journey through space, a non-nominal solar wind as well as turbulent fluctuations may strongly affect the SEP trajectories, altering both the spatial and temporal dependencies of energetic particle distributions in the heliosphere.
Non-nominal solar wind conditions may for example be caused by interplanetary CMEs, or by solar wind streams of different speeds interacting with each other. 
When a fast solar wind stream is followed by a slow stream, a rarefaction region forms between the two, whereas in the opposite case, the high-speed stream compresses the slow solar wind, which may eventually result in the formation of a corotating interaction region (CIR) \cite{smith76}. 
CIRs may accelerate particles at large heliocentric radial distances, where one or both of the shock waves bounding the CIR are present, via diffusive shock acceleration \cite{bell78}, and at smaller radial distances due to the compressional waves bounding the transition region between the slow and the fast solar winds, e.g., \cite{desai98,giacalone02}.
 
To study the influence of a non-nominal IMF on SEP events, Wijsen et al.~\cite{wijsen19a} and \cite{wijsen19b} used a three-dimensional particle transport model to propagate particles in a solar wind generated by the EUropean Heliospheric FORecasting Information Asset (EUHFORIA) \cite{pomoell18}, which is a three-dimensional magnetohydrodynamic (MHD) model of the solar wind and CME evolution in the inner heliosphere. 
Using EUHFORIA, these authors modelled a fast solar wind stream embedded in a slow solar wind, generating self-consistently a CIR from ${\sim} 1.3 $ AU onward. 
By considering impulsive injections of 4 MeV protons from different extended injection regions located at 0.1 AU, Wijsen et al.~\cite{wijsen19b} illustrated how the longitudinal width of the particle intensity distributions can strongly increase or decrease as a function of radial distance, depending on the underlying IMF topology.  
Moreover, by performing simulations both with and without cross-field diffusion, it was shown that  cross-field diffusion widens the region where particles propagate, i.e., the particle streaming zone, yet it does not influence the qualitative 
behaviour of the longitudinal width as a function of radial distance \cite{wijsen19b}.  
We want to point out that we leave it for future work to study the three-dimensional spread of particles in the early phase of SEP events, that is, where particles propagate closer to the Sun than our inner boundary at 0.1 AU. 
Both observational and modelling studies indicate that particles may be accelerated by shocks associated with CMEs already $< 10$\,R$_{\odot}$ (e.g., \cite{roussev04,lee12,rouillard12,pomoell15,lario16,schwadron17,afanasiev18,kozarev19}, and references therein). 
In addition, shock waves associated with CMEs contribute to the spread of SEPs in interplanetary space (e.g., \cite{cane06}). 
Recently, Hu et al.~\cite{hu18} used the two-dimensional iPATH model (\cite{hu17}) to model both diffusive shock acceleration and interplanetary transport of protons starting from $10$\,R$_{\odot}$ during a gradual SEP event, as seen by virtual observers located at different radial distances in the solar equatorial plane, and covering a $60^\circ$ azimuthal range. 
Assuming nominal solar wind conditions upstream of the CME driven shock wave, these authors used the Nonlinear Guiding Center (NLGC) theory \cite{matthaeus03,zank04,shalchi09} to  illustrate  how cross-field diffusion can alter the intensity-time profiles for 4.4~MeV and 46.4~MeV particles, especially for observers magnetically not well-connected with the travelling shock. 
The intensity profiles measured by these observers showed more prompt enhancements for simulations with larger cross-field diffusion coefficients.

In this work, we extend the studies performed in \cite{wijsen19a} and \cite{wijsen19b} by applying different cross-field diffusion conditions to the particles propagating in non-nominal solar wind conditions in the interplanetary space.
In particular, we investigate how a larger mean free path perpendicular to the magnetic field might alter the results obtained in these previous studies. 
After briefly introducing our energetic particle and solar wind models in  Section~\ref{sec:model}, we show how the longitudinal width of the particle intensity distribution varies as a function of radial distance, time, and level of cross-field diffusion (Section~\ref{sec:particle_spread}). 
Next, we investigate how the different levels of cross-field diffusion used in our simulations affect the particle acceleration at the shock waves bounding the CIR (Section~\ref{sec:CD_acc}).
We end in Section~\ref{sec:discussion} with a summary and discussion of the obtained results.

\section{Simulation of SEP transport within a CIR}\label{sec:model}
To study SEPs in the inner heliosphere, we model the evolution of the directional particle intensity
$j(\vec{x},p,\mu,t)$
using the focused transport equation (FTE) (see e.g. \cite{roelof69, isenberg97,leRoux09}):
\begin{equation}\label{eq:fte}
\begin{aligned}
\pd{j}{t} &+\pd{}{\vec{x}}\cdot\left[\left(\td{\vec{x}}{t}+\pd{}{\vec{x}}\cdot \vec{\kappa}_\perp\right)j\right]+\pd{}{\mu}\left[\left(\td{\mu}{t}+ \pd{D_{\mu\mu}}{\mu}\right) j\right] + \pd{}{p}\left(\td{p}{t}j\right)\\ &= \pdd{}{\mu}\left[D_{\mu\mu}j\right] + \pd{}{\vec{x}}\cdot\left[\pd{}{\vec{x}}\cdot\left(\bm{\kappa}_\perp j\right)\right],
\end{aligned}
\end{equation}
with
\begin{eqnarray}
\td{\vec{x}}{t} &=& \vec{V}_{\rm sw}+ \mu \varv \vec{b} +\vec{V}_d  \label{eq:fte_x}  \\
\td{\mu}{t}&=&\frac{1-{\mu}^2}{2}\Bigg(\varv\nabla\cdot\vec{b}+ \mu \nabla\cdot\vec{V}_{\rm sw}- 3 {\mu} \vec{b}\vec{b}:\nabla\vec{V}_{\rm sw} - \frac{2}{{\varv}}\vec{b}\cdot\td{\vec{V}_{\rm sw}}{t}\Bigg)   \label{eq:fte_mu} \\
\td{p}{t} &=& \Bigg(\frac{1-3{\mu}^2}{2}(\vec{b}\vec{b}:\nabla\vec{V}_{\rm sw}) - \frac{1-{\mu}^2}{2}\nabla\cdot\vec{V}_{\rm sw} -\frac{{\mu} }{{\varv}}\vec{b}\cdot\td{\vec{V}_{\rm sw}}{t}\Bigg){p} \label{eq:fte_p}.
\end{eqnarray}
In these equations, 
$\vec{x}$ denotes the phase-space spatial coordinate and $t$ the time, both measured in an inertial frame, whereas the cosine of the pitch angle $\mu$ and the momentum magnitude $p$ or speed $\varv$ are expressed in a frame co-moving with the solar wind. 
Furthermore, $\vec{V}_{\rm sw}$ is the solar wind velocity, and
$\vec{b}$ the unit vector in the direction of the mean magnetic field,
$D_{\mu\mu}$ is the pitch-angle diffusion coefficient, 
$\bm{\kappa}_\perp$ the spatial cross-field diffusion tensor, and
$\vec{V}_d$ the drift velocity due to the gradient and curvature of the mean magnetic field.  
Our particle transport model solves Eq.~\eqref{eq:fte} by integrating the equivalent set of It\^o stochastic differential equations forward in time \cite{gardiner04}.  
The solar wind velocity and magnetic field variables appearing in Eqs.~\eqref{eq:fte_x}--\eqref{eq:fte_p} are obtained from the three-dimensional MHD model EUHFORIA. For details regarding the numerical procedures used by our model, we refer to \cite{wijsen19a}. 

As in \cite{wijsen19a} and \cite{wijsen19b}, we use EUHFORIA to model a slow solar wind with an embedded fast solar wind stream. 
This is done by prescribing everywhere at the inner boundary of EUHFORIA a solar wind speed of 330~km/s, except for a region satisfying
 $
(\rm{longitude}-75^\circ)^2 + (\rm{latitude}-5^\circ)^2< (20^\circ)^2,
$
where we prescribe a solar wind of speed 660~km/s. 
Moreover, this high-speed region is surrounded by a transition zone of width $6^\circ$. 
This transition eventually evolves into a CIR, bounded by reverse and forward MHD shock waves. 
As discussed in \cite{wijsen19a}, the precise location where the large-amplitude compression waves bounding the transition region evolve into shock waves, and hence marking the formation of the CIR, is difficult to pinpoint. 
This is because the finite resolution of our MHD simulation produces inevitably shock waves that are wider than observed interplanetary CIR shocks. 
However, we observe in our particle simulations that particles start to get accelerated from a heliographic radial distance of ${\sim}1.3$ AU onward, indicating that the large-amplitude compression waves have steepened sufficiently such that their characteristic length scale is considerably smaller than the particle mean free path that we assume (see below). 

In contrast to the EUHFORIA simulation used in \cite{wijsen19a} and \cite{wijsen19b}, we assume in the current study a monopolar positive magnetic field at the inner boundary of the model, such that there is no current sheet present in our solar wind. 
This is done to avoid any effects of the current sheet on the simulated particle intensities, which is necessary due to the larger spread of the particles in this work compared to the simulations performed in \cite{wijsen19a} and \cite{wijsen19b}.
We want to note that 
the guiding centre approximation, used in Eq~\eqref{eq:fte}, breaks down near the current sheet \cite{speiser65}. 
Apart from the magnetic field polarity, all other MHD variables prescribed at the inner boundary of the model are the same as in 
\cite{wijsen19a}, and hence we refer to that paper for further details. 

The simulated source of particles is located at the inner boundary of the model, at a heliographic radial distance of 0.1 AU, and centred around the equatorial plane with a latitudinal width of $10^\circ$. 
The source has a longitudinal width of $30^\circ$ such that it covers a region extending from the slow to the fast solar wind.   
As a consequence, the particle injection region is magnetically connected to both the forward and reverse compression/shock waves, hence also  encompassing the SI inside the CIR, where the shocked slow and fast solar wind plasma meet.
We note that the injection region considered in this work is identical to the one of \cite{wijsen19a} and to case~4 of \cite{wijsen19b} (see Fig~1 of \cite{wijsen19b}).  
Likewise, we also consider a mono-energetic injection of 4~MeV protons, allowing us to easily track the energy changes of the particles during their propagation in the solar wind.
Unlike in \cite{wijsen19a} and \cite{wijsen19b}, in order to mimic a prompt injection of particles, we inject the protons according to a Reid-Axford time profile \cite{reid64}:
\[
j(t) \propto \frac{1}{t}\exp{\left(-\frac{\beta}{t} - \frac{t}{\tau}\right)},
\]
where we choose the parameters $\beta = 1.5 $~hours and $\tau=0.2$~hours (like in \cite{lario07}).

The particle scattering conditions are determined by the pitch-angle diffusion coefficient $D_{\mu\mu}$ and the cross field diffusion tensor $\bm{\kappa}_\perp$. 
As detailed in  \cite{wijsen19a}, we use quasi-linear theory to prescribe the functional form of $D_{\mu\mu}$.
The strength of the magnetic turbulence acting as pitch-angle scattering centres is quantified by assuming a constant radial mean free path $\lambda_r^\parallel = 0.3$ AU for 4~MeV protons. 
 For the cross-field diffusion tensor we assume an expression similar to the one used by e.g. \cite{zhang09,wang12}:
\begin{eqnarray}\label{eq:cross_field}
\bm{\kappa}_\perp &=&\alpha \frac{\pi }{12}{ {\varv}\lambda_\parallel}\frac{B_0}{B}\left(\mathbb{I}-\vec{bb}\right),
\end{eqnarray}
where  $\mathbb{I}$ is the unit tensor, 
$\vec{bb}$ the dyadic product of the magnetic unit vector, and 
$\lambda_\parallel$ the parallel mean free path.
$B_0$ is a reference magnetic field strength value fixed to  the maximum value of $B$ at 1 AU, 
$B_0 = 9.7$~nT, which is attained in the shocked slow solar wind, adjacent to the SI. 

We note that we used rather simple functional forms for the pitch-angle and cross-field diffusion coefficients. 
However, the significant uncertainties regarding the turbulent conditions and hence the particle diffusion properties in large scale structures like CIRs, make it difficult to establish more realistic assumptions for the particle diffusion coefficients \cite{horbury99, crooker99,richardson18}.
Nonetheless, in Wijsen et al.~\cite{wijsen19a,wijsen19b} it is demonstrated how our choice for the pitch-angle and the cross-field diffusion coefficients can reproduce different features of the energetic particle intensity profiles associated with CIRs, like e.g. the particle acceleration near the shock waves and the intensity dip near the SI. 
This is also discussed in more detail in Section~\ref{sec:discussion}. 
We remark that in reality the parallel and perpendicular mean free path of the particles might vary rather strongly across the CIR due to the presence of different turbulent regimes \cite{horbury99}. 
The increase in turbulence near the CIR shocks (see  e.g, \cite{tsurutani95} and Section~\ref{sec:discussion} for a discussion) would most likely decrease the parallel mean free path and amplify some of the effects that we describe in this work. 
A smaller radial mean free path would e.g., imply that the particles will scatter across the CIR shock waves more often, resulting in more particle acceleration. 
On the other hand, using a large constant radial mean free path everywhere in our simulations allows us to more easily trace the effects of the underlying non-nominal IMF topology and the different levels of cross-field diffusion. 
In addition, we note that more advanced theories of the cross-field diffusion, like e.g., NLGC theory, require the specification of the magnetic turbulence level $\delta B$ in the heliosphere. 
For nominal solar wind conditions, one typically assumes that this quantity decreases as a power-law with radial distance from the Sun (see e.g., \cite{mason12,zhao16, hu17}). 
However, such a singular dependence on the radial coordinate is not valid in a solar wind containing a CIR, as the turbulent conditions in a CIR are significantly different from those encountered in the fast or slow solar wind (see e.g., \cite{horbury99, crooker99,richardson18}). 
Therefore, since the behaviour of $\delta B$ in a CIR is not well understood, we instead adopt the phenomenological form~\eqref{eq:cross_field} for the cross-field diffusion, as its effects are more easily traceable in the simulations.

By defining the perpendicular mean free path as $\lambda_\perp = 3\kappa_\perp/\varv$, the parameter $\alpha$ determines the ratio $\lambda_\perp/\lambda_\parallel$ at 1 AU up to a factor of order unity.
We consider three different values for the parameter $\alpha$, namely $2\times10^{-4}$, $2\times10^{-3}$, and $2\times10^{-2}$. In the remainder of the text, we will refer to the simulations performed with these different values of $\alpha$ as case~1, case~2 and case~3, respectively. 
These values of $\alpha$ help us to classify the `level' or `strength' of the cross-field diffusion employed in our simulations. 
In this way, the case with the strongest cross-field diffusion corresponds to case~3 ($\alpha = 2\times10^{-2}$).

\section{Evolution of the particle spread in the solar equatorial plane}\label{sec:particle_spread}
\begin{figure*}
        \centering
        \begin{tabular}{cc}
        \includegraphics[width=0.45\textwidth]{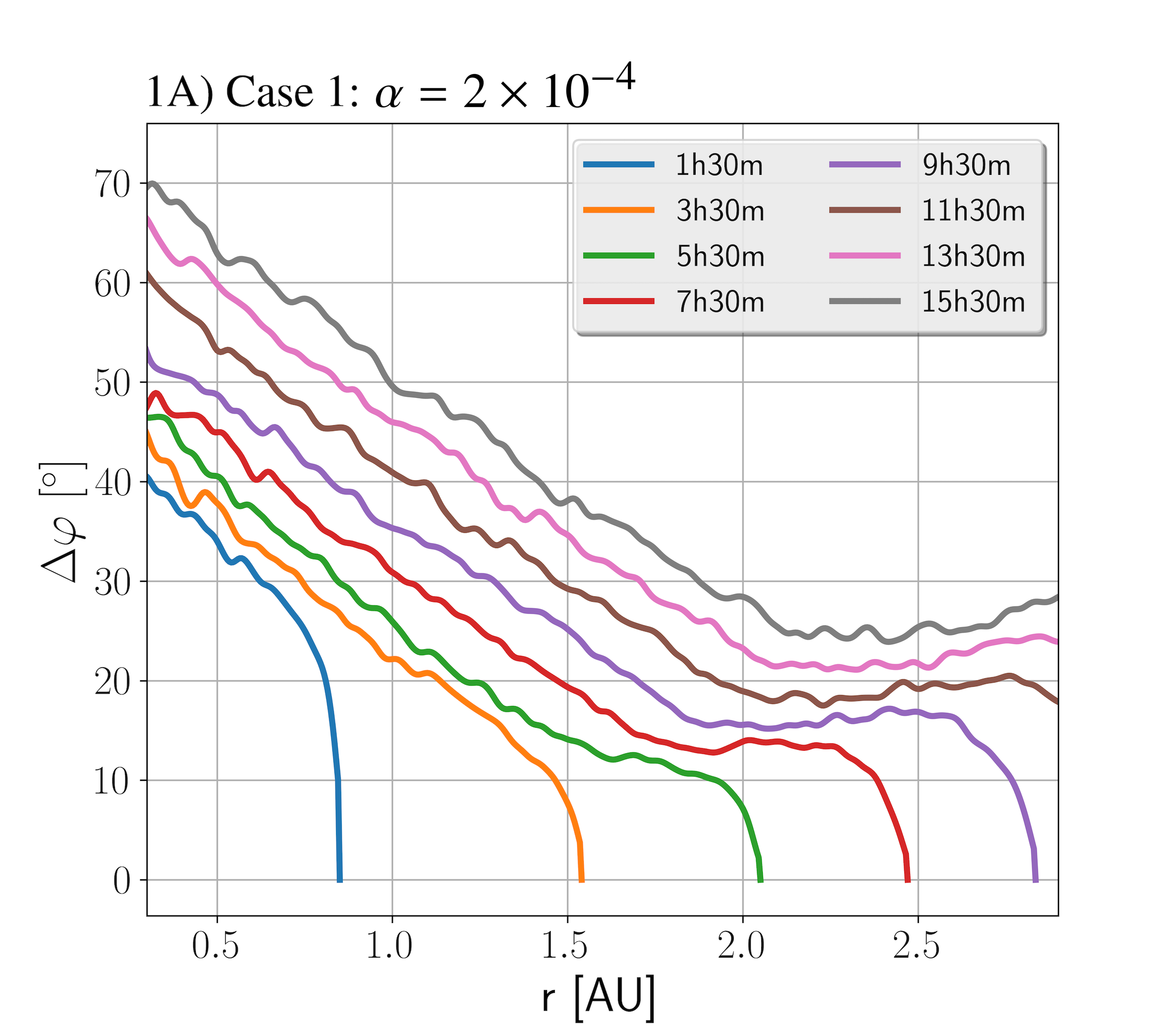}
        &
        \includegraphics[width=0.45\textwidth]{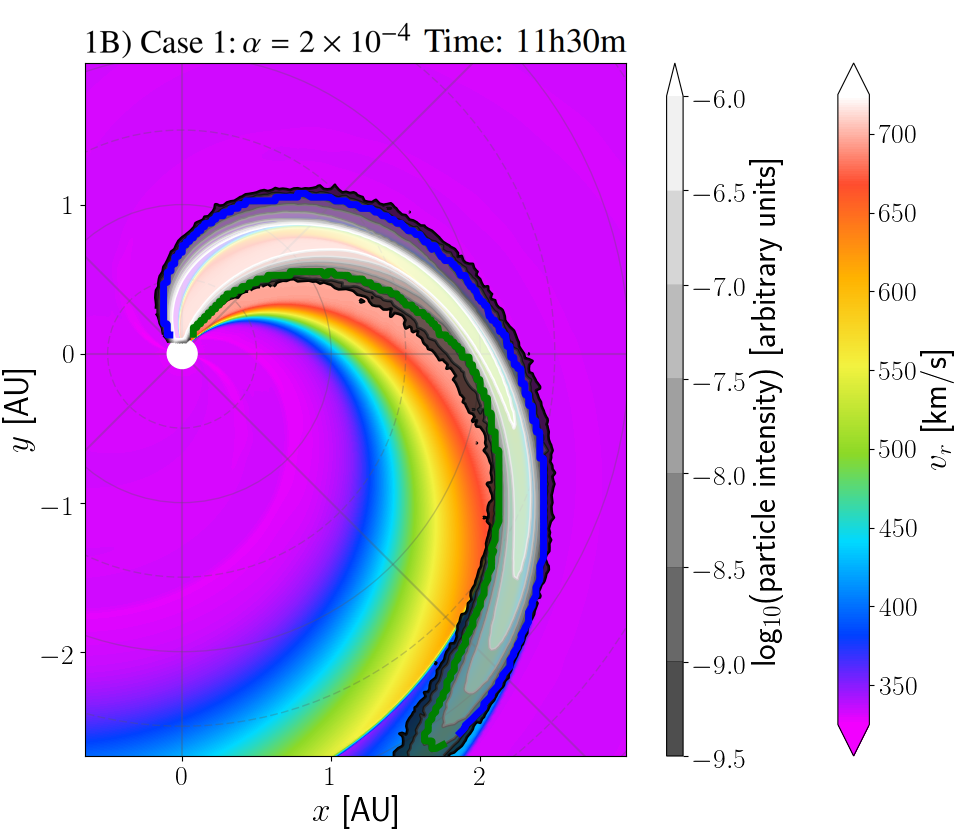}\\
        \includegraphics[width=0.45\textwidth]{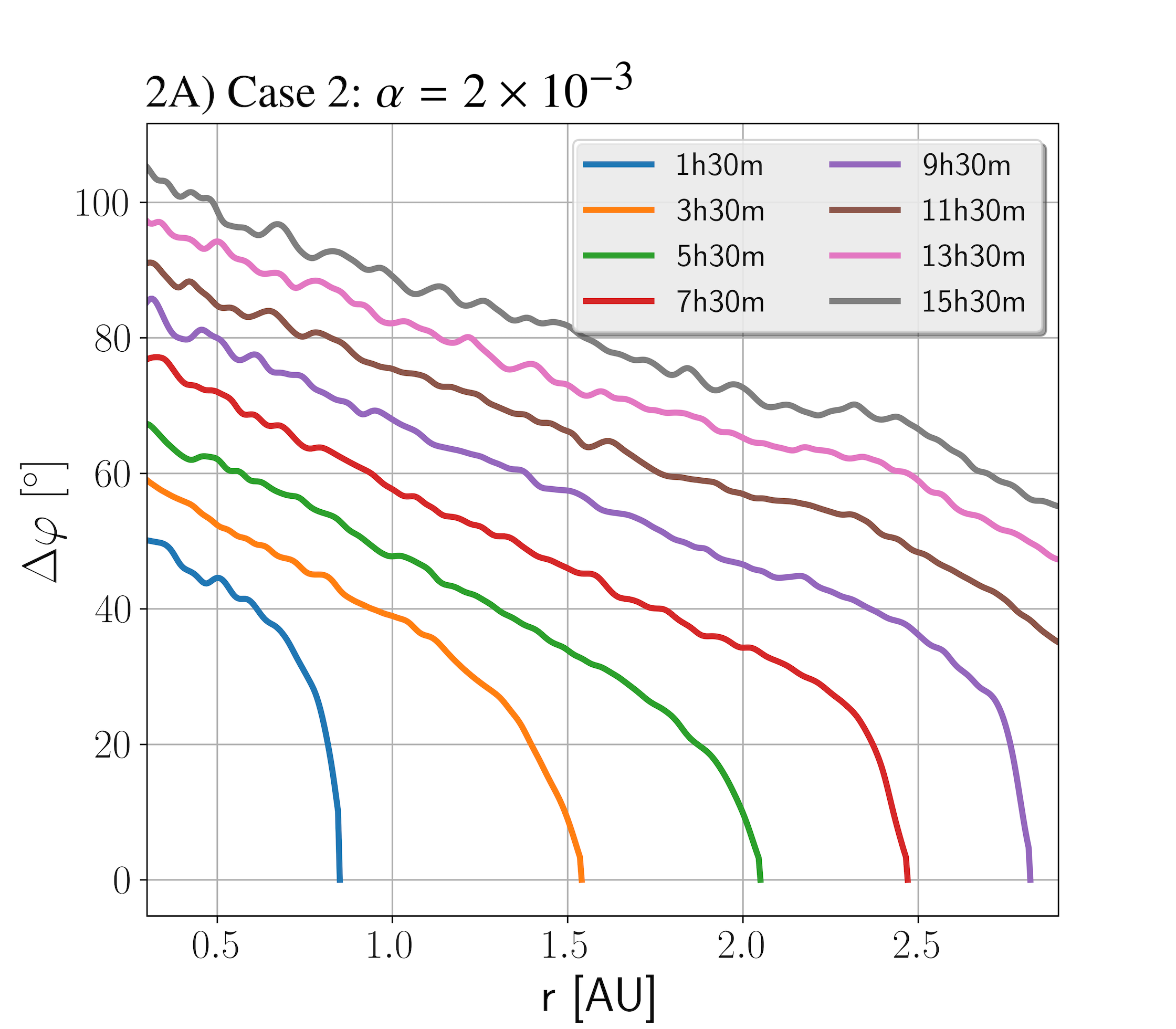}&
        \includegraphics[width=0.45\textwidth]{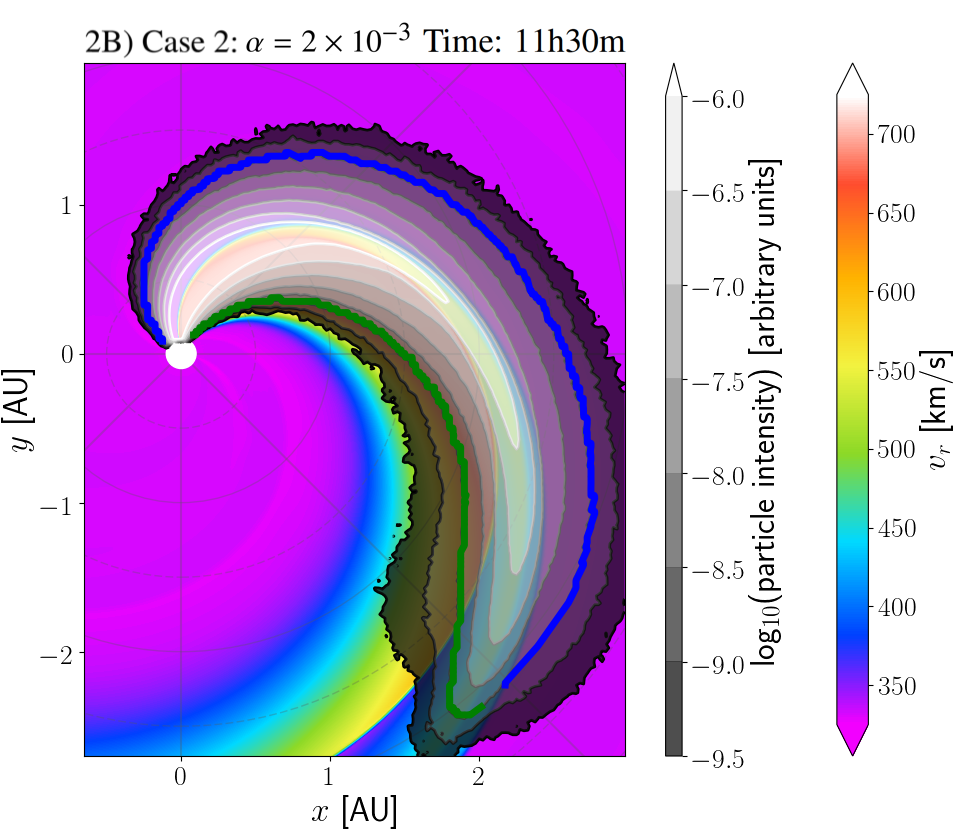}\\
        \includegraphics[width=0.45\textwidth]{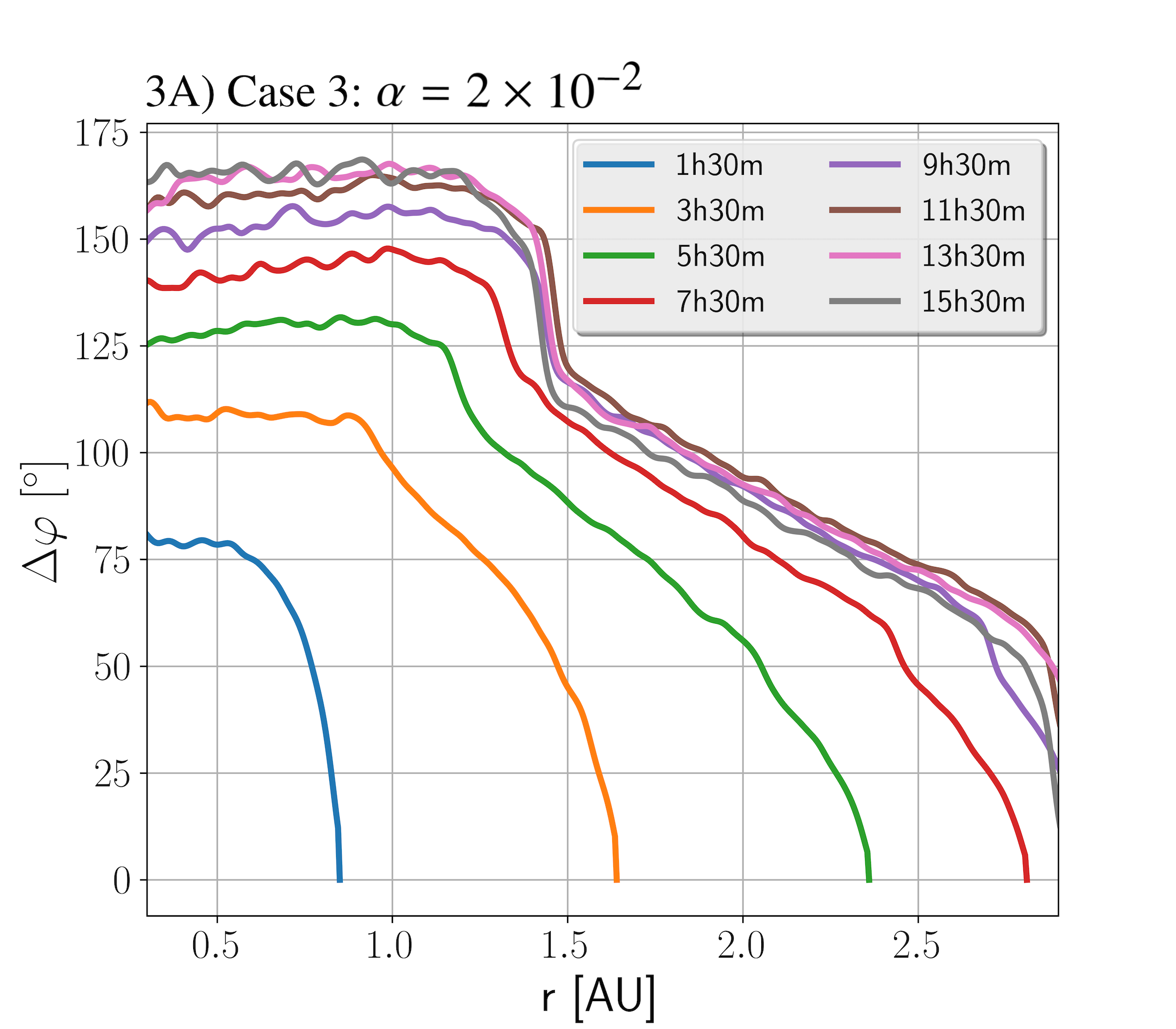}&
        \includegraphics[width=0.45\textwidth]{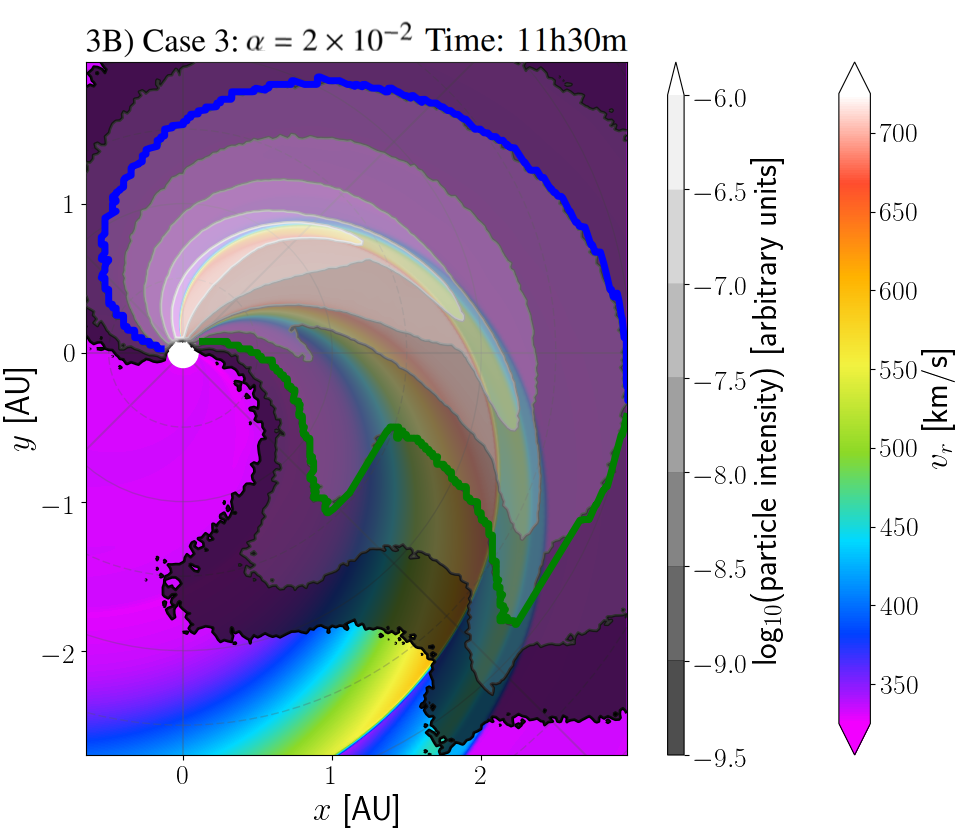}\\
    \end{tabular}
    \caption{Column~A (left column) shows the longitudinal width of the omni-directional intensity distribution as a function of radial distance for the different indicated times.  
    Column~B (right column) shows a snapshot at 11.5~hours of the omni-directional particle intensities, drawn in grey shades on top of the radial velocity profile of the solar wind. 
    The green and blue curves represent, respectively, the lower and upper longitudinal boundaries of the streaming zone, as determined by the prescribed intensity threshold value $10^{-8.5}$ (arbitrary units). 
 The first row is for case~1, the second row for case~2, and the third row for case~3. }
     \label{fig:dlon}
\end{figure*}
In this section, we analyse how the different levels of cross-field diffusion affect the longitudinal width of the particle intensity distributions in the solar equatorial plane.
For this purpose, we define the particle streaming zone as the region where the omni-directional intensity, integrated over energy, is larger than the limiting value $10^{-8.5}$ (in arbitrary units). 
This threshold is set three orders of magnitude below the peak-intensity attained at 1 AU for a simulation without cross-field diffusion. 
Defining the longitudinal coordinate $\varphi$ in anticlockwise 
direction, we obtain at each heliographic radial distance, $r$, and time, $t$, the lower, $\varphi_L(t,r)$, and upper, $\varphi_U(t,r)$, edge longitudes of the particle streaming zone, such that we can define the longitudinal width of the particle streaming zone as $\Delta\varphi(r,t) =\varphi_U(t,r) - \varphi_L(t,r)$. 
The left column of  Fig.~\ref{fig:dlon} shows for the three cases the radial dependence of $\Delta\varphi(r,t)$ at different times. 
The right column of Fig.~\ref{fig:dlon} shows for each case the corresponding snapshots of the particle intensities at 11.5~hours. 
In these figures, the lower longitudinal boundary is indicated by a green line, whereas the upper longitudinal boundary is  shown in blue. 

From the first column of Fig.~\ref{fig:dlon}, we see that for all cases, $\Delta\varphi(r,t)$ generally decreases with increasing radial distance, reflecting the IMF compression at the shock waves bounding the CIR (see also Fig.~6 of \cite{wijsen19a}).
However, apart from this common decreasing trend, there are several differences between the radial evolution of $\Delta\phi$ when comparing the different cases.  

For case~1, panel~1A shows that at large radial distances ($r\gtrsim 2$ AU), the width of the particle streaming zone remains approximately constant and even starts to increase slightly.
At these distances, the particle streaming zone is located entirely within the CIR, meaning that all the IMF lines on which particles are travelling have crossed the shock waves at distances $r<2$ AU. 
In contrast, we see in panel~2A that for case~2 the longitudinal width of the particle streaming zone does not become constant from $r=2$ AU onward. 
This is because the stronger cross-field diffusion has brought particles to IMF lines that cross the shock waves at radial distances larger than $2$ AU, and hence, contrary to case~1, there are still particles outside the CIR at these large radial distances.
However, since the forward and reverse shock waves propagate into the slow and fast solar wind, respectively, we expect that from a certain radial distance onward, all particles will again reside inside the CIR, and hence the behaviour of the longitudinal width will probably become qualitatively similar to what is observed for case~1. 

Panel~3A shows that the strong cross-field diffusion of case~3 introduces several features to the radial dependence of the longitudinal width that are not seen in the two other cases.  
At small radial distances, the width of the particle streaming zone remains approximately constant, and even increases slightly, which is because some particles have been transported to the unperturbed slow solar wind behind the high-speed stream. 
However, panel 3A shows that the constant longitudinal width of the particle intensity distribution is followed by a sudden drop at larger radial distances.
For times later than 9.5 hours, this drop occurs at approximately the same radial distance, $r=1.4$~AU.
Looking at panel 3B, we see that this decrease is because the omni-directional intensity drops below the $10^{-8.5}$ threshold in the rarefaction region behind the high-speed solar wind stream (the pronounced indent in the green intensity threshold contour in Figure 3B).
This is a result of the strong expansion of magnetic flux tubes inside the rarefaction region, which causes the energetic particle density to get rarefied as well. 
In addition, the large cross-field diffusion for case~3 spreads the energetic particles over a much larger region than for the two other cases, hence rarefying the particle densities everywhere in the heliosphere. 
This is especially true in the fast solar wind stream and the rarefaction region, since the low magnetic field strength implies a more efficient cross-field diffusion (see Eq.~\eqref{eq:cross_field}).
As a consequence, the intensities for case~3 are everywhere relatively low as compared to the previous cases. 
These low intensities might imply that a large part of the SEP event may go undetected if  the pre-event intensity levels are high enough or the measuring instrument is not sensitive enough. 
In fact, such a situation is illustrated in panel~3B, where we see that a large region is characterised by particle intensities below the $10^{-8.5}$ threshold, which is therefore not taken into account when determining the width of the intensity distributions.
The low intensity values due to the efficient particle spreading also explains why in panel~3A the longitudinal width does not increase any further after 9.5~hours, and why for a radial distance larger than 1.5~AU, the longitudinal width at time 15.5~hours is even smaller than at the three previous time instances. 

From the first column of Fig.~\ref{fig:dlon}, we note that for all three cases the longitudinal width of the particle streaming zone at small radial distances and at 1.5~hours is already significantly larger than the longitudinal width of the injection region, which was $30^\circ$.
A rapid spread of particles at early times can be understood by considering the diffusive cross-field flux, which can, for  small radial distances and in the solar equatorial plane, be approximated as in \cite{strauss17b}:
\[
\left|\mathcal{F}_\perp^{\rm dif}\right|\sim \kappa_\perp  \left|\pd{j}{\varphi} \right|.
\]
From this expression, we see that the cross-field diffusion will be more efficient when large longitudinal gradients are present in the directional intensity function. 
Since our particles are injected with a uniform intensity from a source region of limited spatial extend, we have in fact that, upon injection,  $\left|\partial{j}/\partial{\varphi} \right| \longrightarrow +\infty$ at the boundaries of our injection region in the solar equatorial plane. 
We note that this infinite derivative does not produce numerical problems due to the stochastic nature of our model. 
However, it explains the efficient cross-field diffusion early on in the event (see also \cite{strauss17b} for a discussion).

Moreover, we note that the above argument is more generally valid at any location in the heliosphere when considering the gradient of the intensity in the direction perpendicular to the magnetic field. 
More specifically, for an isotropic $\bm{\kappa}_\perp$, the diffusive cross-field flux in the direction of a unit vector $\vec{e}_\perp$, with $\vec{e}_\perp \perp \vec{b}$, can be defined as 
\begin{equation}\label{eq:cross-fieldFlux}
  \left|\mathcal{F}_\perp^{\rm dif}\right|\sim \kappa_\perp  \left|\vec{e}_\perp\cdot\nabla{j}\right|.  
\end{equation}
Hence, cross-field diffusion will be most effective where $\kappa_{\perp}$ is large or where there are large particle intensity gradients perpendicular to the magnetic field. 
Such large gradients are induced by the IMF topology inside the CIR, where the magnetic field is compressed.
As discussed above, this compression of the magnetic field lines leads to a compression of the particle intensity distribution, and hence a decreasing longitudinal width. 
However, the same compression will also produce larger particle intensity gradients perpendicular to the magnetic field, hence increasing the effectiveness of cross-field diffusion according to Eq. \eqref{eq:cross-fieldFlux}. 
At the same time, we note that the inverse scaling of our cross-field diffusion with the magnetic field strength implies that the perpendicular mean free path decreases inside the CIR. 
However, as discussed in \cite{wijsen19a} and \cite{wijsen19b}, a small cross-field motion of the particles inside the CIR where the IMF lines are closely compressed may transport the particles on field lines that are significantly separated outside the  CIR, i.e., at smaller radial distances. 
The observed cross-field diffusion in our simulations results thus from a non-trivial interplay between the magnetic field dependence of the cross-field diffusion tensor and the intensity gradients perpendicular to the magnetic field. 
Both mechanisms are strongly influenced by non-nominal solar wind conditions.

Finally, we remark that by adding the cross-field diffusion term in the spatial part of the FTE, i.e., in Eq.~\eqref{eq:fte_x}, the energetic particles  are given an additional velocity component on top of their actual velocity $\vec{\varv}$.
 As a consequence, particles with $\mu \sim 1$ might spread faster than their speed allows for, hence violating causality  \cite{strauss15}. 
 Following a similar procedure as in~\cite{strauss15} to estimate the impact of this effect in our simulations, we found it to be negligible since the intensities resulting from such particles are more than two orders of magnitude below the selected threshold value. 
This non-physical behaviour is a consequence of treating the motion of particles perpendicular to the mean magnetic field as a spatial diffusion process, and would not appear if one would instead e.g., model cross-field motions by stochastically varying only the propagation direction of the particles. 
Such an approach boils down to  varying the magnetic unit vector, which is similar to the method implemented by Laitinen et al.~\cite{laitinen13} to model the  motion of particles along random-walking magnetic field lines. 

\section{The effect of cross-field diffusion on shock accelerated particle populations}\label{sec:CD_acc}

\begin{figure*}
        \centering
        \begin{tabular}{cc}
        \includegraphics[width=0.45\textwidth]{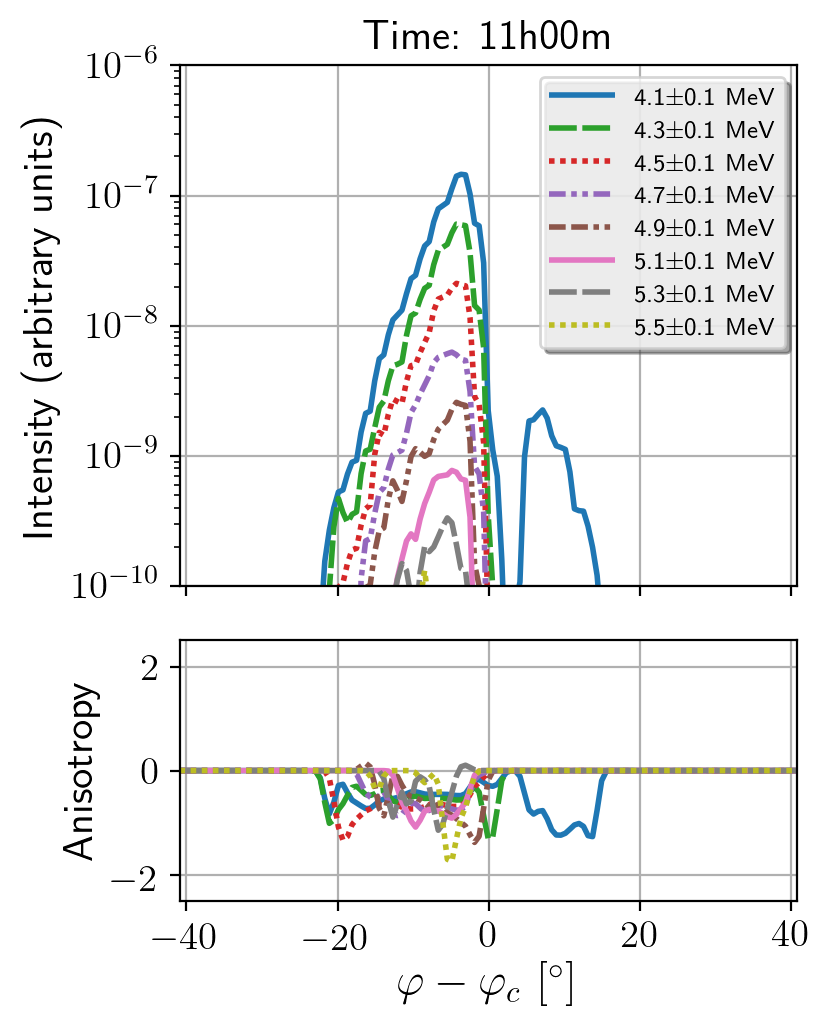}&
        \includegraphics[width=0.45\textwidth]{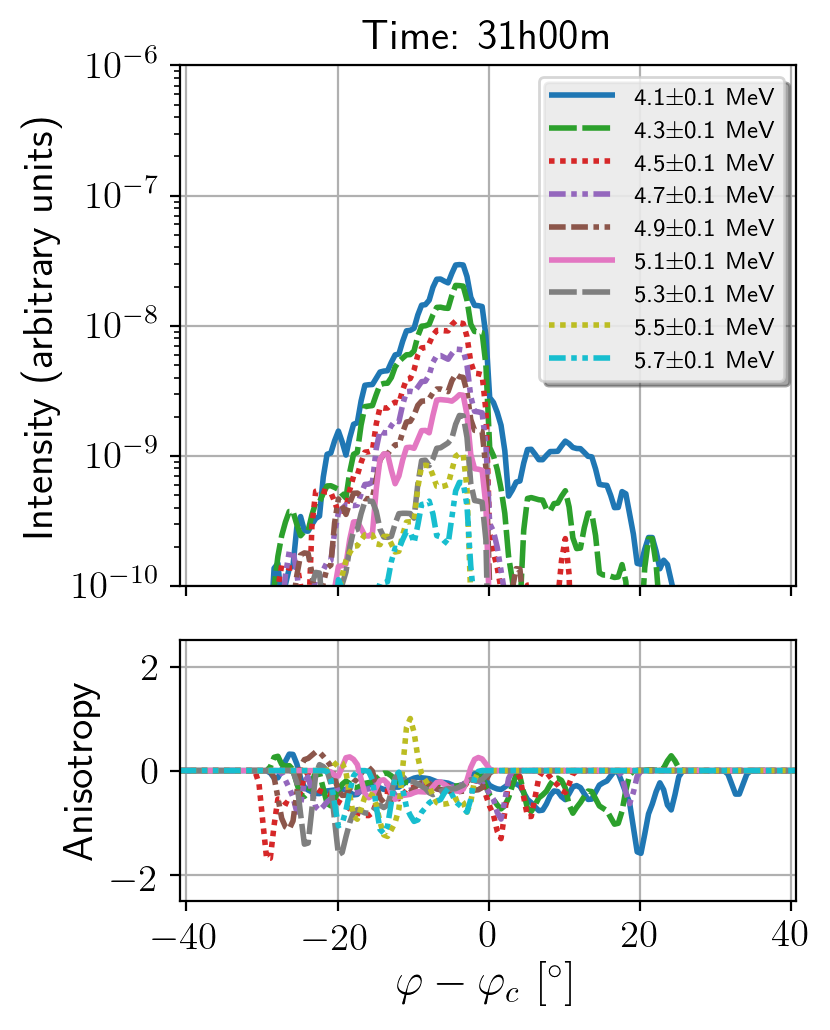}
    \end{tabular}
    \caption{Snapshot of the proton intensities (upper panels) and anisotropies (lower panels) for case~1  at 11.0~hours (left panels) and 31.0~hours (right panels). The profiles are measured in the solar equatorial plane at a fixed heliocentric radial distance of 1 AU, and shown as a function of the longitudinal separation from the centre of the injection region.}
     \label{fig:lon1}
\end{figure*}

\begin{figure*}
        \centering
        \begin{tabular}{cc}
        \includegraphics[width=0.45\textwidth]{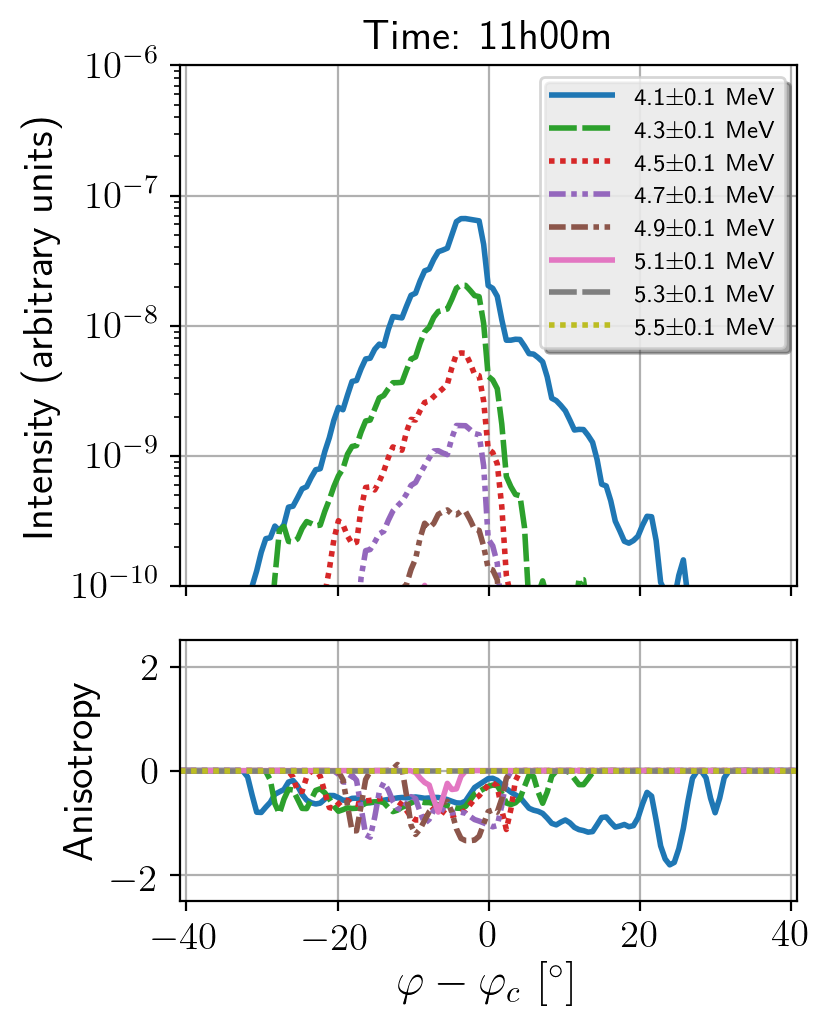}&
        \includegraphics[width=0.45\textwidth]{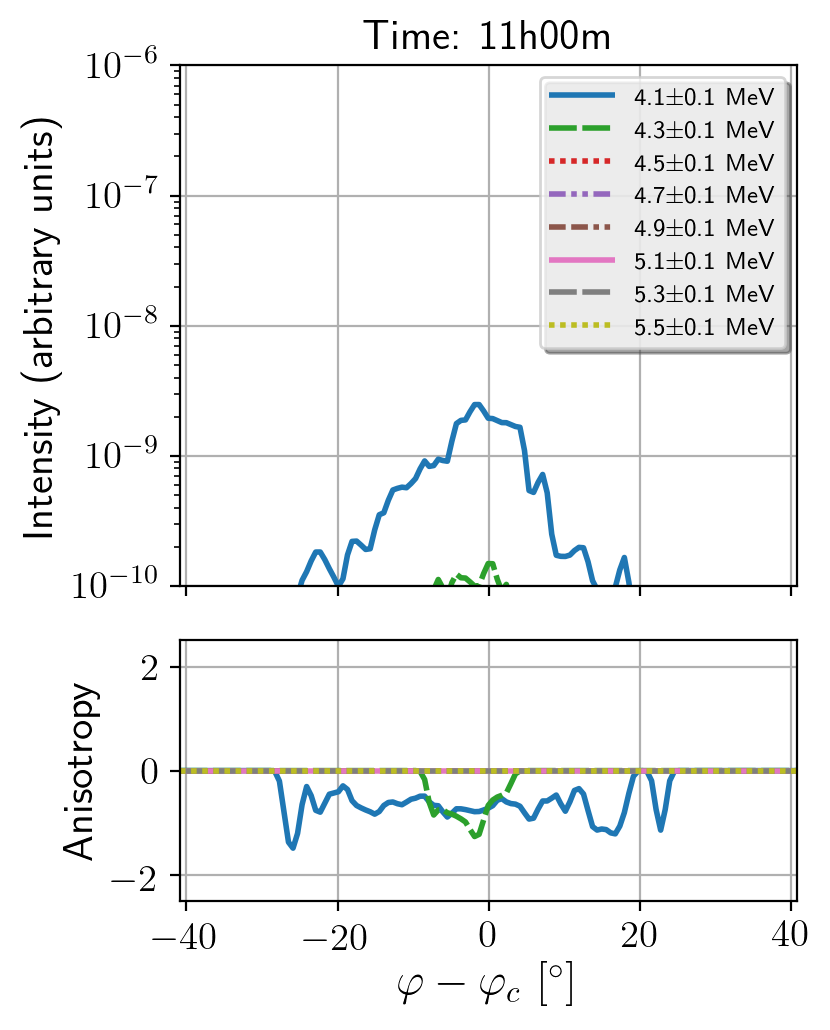}\\
    \end{tabular}
    \caption{Snapshot of the proton intensities (upper panels) and anisotropies (lower panels) at time 11.0~hours  for case~2  (left panel) and case~3  (right panel). The profiles are measured in the solar equatorial plane at a fixed heliocentric radial distance of 1 AU, and shown as a function of the longitudinal separation from the centre of the injection region.}
     \label{fig:lon2}
\end{figure*}

We now examine the particle intensity and anisotropy profiles in the solar equatorial plane at a fixed heliographic radial distance of 1~AU. 
We focus hereby on energy channels above the initial 4~MeV injection energy of the particles, meaning that we only consider particles that have been accelerated at the shock waves bounding the CIR. 
We note that the acceleration of the protons and ions in shock waves depends on several factors, including the scattering conditions (e.g., \cite{lee83,gordon99, vainio07,afanasiev15}), the shock strength (e.g., \cite{lario98,rice03,gasen11,gasen14,prinsloo19}), and the shock obliquity (e.g., \cite{zank01,zank06,battarbee13}), among others. 
In Section~\ref{sec:discussion} we provide a discussion on these factors in the context of CIR shocks, and how they influence the results presented in this section. 

Figure~\ref{fig:lon1} shows two snapshots, at 11.0~hours (left) and at 31.0~hours (right), of the intensity and anisotropy profiles for case~1, as a function of $\varphi - \varphi_c$, where $\varphi_c$ denotes the longitude of the point at $1$ AU that connects magnetically to the centre of the particle injection region at 0.1 AU. 
The left panel of Fig.~\ref{fig:lon1} shows two distinct intensity enhancements in the $4.1\pm 0.1$ MeV energy channel, resulting from particle acceleration at the two shock waves bounding the CIR. 
The intensity enhancement at $\varphi - \varphi_c <0$ is produced by particle acceleration at the reverse shock, whereas the intensity enhancement at $\varphi - \varphi_c >0$ is produced by particle acceleration at the forward shock. 
The negative anisotropies in the lower panels of Fig.~\ref{fig:lon1} indicate that the acceleration of the particles occurs at radial distances larger than 1~AU. 
This is in accordance with the findings of Wijsen et al.~\cite{wijsen19a}, where we observed the generation of accelerated particles at the shocks for $r \gtrsim$ 1.3 AU. 
In our simulations, the reverse shock is a considerably more efficient particle accelerator than the forward shock. 
For the former, we see that after 11~hours there are particles accelerated up to ${\sim}1.0$~MeV above their initial energy, whereas the latter only accelerates particles slightly above 4~MeV.

The two intensity enhancements in the left panel of Fig.~\ref{fig:lon1} can co-exist without merging partly because the SI separating them is characterised by a strong magnetic field, and hence by a weak cross-field diffusion since the latter scales inversely proportional to the magnetic field strength (see Eq.\eqref{eq:cross_field}). 
However, since the cross-field diffusion is not exactly zero at the SI, the separation between the two intensity enhancements will slowly diffuse away. 
The latter might be aided by the large particle intensity gradients perpendicular to the magnetic field observed at the SI.
Looking at the right panel of Fig.~\ref{fig:lon1}, which shows the intensity and anisotropy profiles of case~1 at 31.0~hours, we realise that the intensity dip has indeed increased to a higher intensity level. By this time, the reverse shock has accelerated protons above 5.3~MeV, although the intensities and hence the statistics for these particles are low.

Figure.~\ref{fig:lon2} shows the longitudinal intensity and anisotropy profiles at 1 AU for case~2 and~3 at 11.0~hours.  
Focusing on the left  panel of Fig.~\ref{fig:lon2}, we see that the stronger cross-field diffusion of case~2 has merged the two $4.1\pm 0.1$ MeV intensity enhancements of case~1 into one wide enhancement.  
Moreover, we see that there is less particle acceleration when compared to the previous case, since the
$5.1\pm 0.1$ MeV and $5.3\pm 0.1$ MeV energy channels are no longer populated by particles. 
The peak intensities of the other channels have also decreased, yet we note that these channels cover now a larger longitudinal range. 
The drop in accelerated particle intensities with increasing cross-field diffusion is even more notable when looking at the right panel of Fig.~\ref{fig:lon2}, which shows the intensity profiles for case~3. 
This panel illustrates that the shock waves are barely capable of accelerating particles due to the large cross-field diffusion assumed in case~3.  
This is because the cross-field diffusion efficiently moves particles away from the shock waves by transporting the particles to IMF lines that either cross the shock wave at much larger radial distances, or that enter the CIR before the shock waves have formed.

\section{Summary and discussion}\label{sec:discussion}

In this work, we have extended the studies performed in \cite{wijsen19a} and \cite{wijsen19b} that investigate SEP transport near and within a high-speed solar wind stream that eventually evolves into a CIR.
For this, we used a three dimensional particle transport model that computes particle directional intensities and anisotropies in a solar wind generated by EUHFORIA.   
In particular, we considered an injection of 4~MeV protons at the inner boundary of our model, from a source region that extends both in the slow and fast solar wind regimes, hence also covering the transition region between both plasma environments.  
Assuming a cross-field diffusion coefficient that scales inversely proportional to the magnetic field strength, we performed simulations with three different levels of cross-field diffusion. 
This allowed us to study how the strength of cross-field diffusion may alter the particles spread and the energy content of SEPs in a solar wind that deviates from a nominal solar wind configuration, under the assumptions of our model.

From our simulations, we found that the longitudinal widths of the particle intensity distributions show a general decreasing trend with increasing radial distance, reflecting the underlying IMF topology near and within the CIR. 
Such a decreasing width is in contrast with what is observed in nominal solar wind configurations, where the width of the particle intensity distribution remains constant in the absence of cross-field diffusion or increases with radial distance if cross-field diffusion is present (see \cite{wijsen19b}). 
Although our three simulations with different levels of cross-field diffusion showed overall a similar decreasing trend, there were some notable differences.  
For the case with the strongest cross-field diffusion, there was a sudden drop in the longitudinal width of the particle intensity distributions at ${\sim}1.4$~AU, because the particle intensity falls below our prescribed threshold. 
This intensity drop occurred in the rarefaction region behind the high-speed solar wind stream, where the magnetic flux tubes are strongly expanding, hence rarefying the SEP densities.
In addition, the strong cross-field diffusion in our simulation dispersed particles over such a large region in the heliosphere that the intensities decreased everywhere significantly in comparison with the simulations that use a weaker cross-field diffusion. 
As a consequence, a substantial amount of particles resided in regions of the solar wind showing intensities below our prescribed threshold, and were therefore not taken into account when calculating the longitudinal width. 
A similar situation may happen with actual in-situ observed SEP events, i.e., when particles propagate either under strong diffusive transport conditions or in solar wind rarefaction regions, they might be dispersed efficiently such that their intensities drop below the background level measured by a given particle instrument, as similarly noticed by Pacheco et al.~\cite{pacheco19} for electron events simulated under diffusive parallel transport conditions.  

Similar to what Strauss et al.~\cite{strauss17b} obtained for electron propagation in a nominal IMF, we find that the cross-field diffusion is very effective early in the event, when large intensity gradients are present. 
However, in contrast to a Parker solar wind, the compressed IMF topology inside the CIR also gives rise to large intensity gradients perpendicular to the magnetic field, increasing the effectiveness of cross-field diffusion. 
In contrast, the strongly diverging magnetic field topology in the rarefaction region might produce very low particle intensity gradients, hence decreasing the diffusive cross-field flux as defined in Eq.~\eqref{eq:cross-fieldFlux}. 
We find thus that for a cross-field diffusion coefficient that scales inversely proportional to the magnetic field strength, the effectiveness of the cross-field diffusion is determined by an intricate interplay between the variations in the magnetic field strength and the particle intensity gradients, both modulated by the IMF topology and hence by the converging or diverging properties of the solar wind flow.  

In Section~\ref{sec:CD_acc}, we showed the intensity profiles of the shock-accelerated particles at 1~AU, as a function of longitude.  
For case~1, i.e., the case with the weakest cross-field diffusion, two separate intensity enhancements were visible, reflecting the particle acceleration occurring at the two different shock waves. 
The negative anisotropies indicate that the acceleration of the particles happens at larger radial distances, where the shocks have formed.  
This agrees with in-situ observations at 1 AU, showing that the flow of energetic ions accelerated at a CIR shock is typically towards the Sun \cite{marshall78}. 

Moreover, the double structure in the intensity profiles is a feature that is also commonly detected with in-situ observations at different radial distances from the Sun \cite{barnes76,decker81,tsurutani82,dwyer97,mason99}. 
Like in our simulations, observations measure typically more accelerated particle fluxes originating from the reverse shock than from the forward shock \cite{tsurutani82,desai98,mason99b}. 
This difference might be partly due to the fact that the relative speed difference between the unperturbed solar wind and the shock is higher for the reverse shock than for the forward shock, since the reverse shock propagates radially inwards in the solar wind frame (see also \cite{03kocharov}). 

Another important factor that might contribute to the observed differences in  accelerated particle fluxes originating from the different shocks is the angle between the shock normal and the upstream magnetic field, i.e. the shock angle, $\theta_{Bn}$. 
In our simulation, the reverse shock of the CIR is more oblique than the forward shock, which follows from the difference in speed between the fast and slow solar wind. 
Tsurutani et al.~\cite{tsurutani82} demonstrated that  proton fluxes near CIRs 
are correlated with the shock angle, with the most intense proton events occurring for shocks that are quasi-perpendicular. 
This is in accordance with the results obtained by Zank et al.~\cite{zank06}, which illustrated that perpendicular shock can be very efficient particle accelerators given that there is a sufficient amount of turbulence present in the solar wind to diffuse the particle across magnetic field lines.
Interesting to note is that CIR shocks are indeed typically characterised by increased amounts of turbulence (see e.g. \cite{horbury99}) as compared to the unperturbed solar wind, despite the fact that wave excitation by the streaming instability is quenched for highly oblique shocks.
The latter is a consequence of the proportionality of the  wave growth to the shock angle \cite{gordon99,li03,rice03}. 

The dip between the two intensity enhancements at the reverse and forward shocks coincides with the SI, 
and can be explained by noting that the field lines adjacent to the SI do not cross the reverse shock wave, nor the forward shock wave. 
Hence the only way for shock-accelerated particles to reach these field-lines is through cross-field transport. 
Using observations made by the Pioneer~10 and~11 spacecraft of a CIR lasting for 14 rotations, Tsurutani et al.~\cite{tsurutani82} showed that the intensity dip increases with a factor of 10 over 11 solar rotations, likely due to cross-field diffusion. 
Similarly, we see in the simulations of case~1 that the intensity minimum increases in time. 
Figure~\ref{fig:lon1} illustrated that the dip increased by a similar factor of ${\sim}10$, yet in only 20 hours, i.e., much faster than the observations in \cite{tsurutani82}. 
This difference can have various causes, as discussed in the following. 
It may suggest that the cross-field diffusion used in case~1, for which $\lambda_\perp/\lambda_\parallel{\sim} 10^{-4}$ at 1 AU, might still be too strong inside a CIR. 
However, Dwyer et al.~\cite{dwyer97} obtained values for $\lambda_\perp/\lambda_\parallel$ of the order of unity for three CIRs using 44\,--\,313~keV/nucleon helium measurements from the Wind spacecraft. 
Important to note here is that the results of Tsurutani et al. \cite{tsurutani82} regarding the increase in the intensity dip near the SI were done when the Pioneer~10 and~11 spacecraft were at a radial distance of ${\sim} 4.4$ and $4.9$ AU, whereas  the Wind spacecraft is located at ${\sim} 1 $ AU, and hence the properties of the solar wind turbulence might be different. 
Other non-negligible factors that affect the acceleration of particles and hence also the evolution of the intensity dip are the plasma normal velocity jump and the compression ratio across the shocks as well as the obliquity of the shocks. 
In fact, Desai et al.~\cite{desai98} showed that the proton intensity at CIRs is well correlated with the compression ratio, even if there is no shock present, at least when the particle seed remains constant. 
In our simulations the particles accelerated at the CIR shocks originate from a flare-like impulsive event, with the particle source region magnetically connected to a limited part of the CIR shocks.  
However, studies have shown that the seed particle population of most CIR events is likely the supra-thermal tail of the solar wind (see e.g. \cite{chotoo00,mason12b}). 
This means that the seed population consists of charged particles with energies around $\sim 10$ keV/nucleon, i.e., much lower than the 4 MeV injection energy that we consider. 
Pitch-angle scattering, cross-field diffusion and the acceleration processes in these non-nominal solar wind conditions are dependent of the momentum of the particles.
A continuous reservoir of low-energy seed particles at the CIR shocks may thus also help explaining the differences observed between our model and the observations in \cite{tsurutani82}. 
However, the observation that CIRs accompanied by the largest energetic particle enhancements are often associated with quasi-perpendicular shocks~\cite{tsurutani82}, raises the question how low energy protons would overcome the associated injection problem \cite{jokipii87,zank96}. 
Possible explanations can be found in the high levels of turbulence found near CIR shocks, or by considering a seed population originating from the supra-thermal tail  of a shock-heated solar wind plasma. 
Prinsloo et al.~\cite{prinsloo19} illustrated that the  prior heating of the solar wind  distribution is important in reproducing the intensities of typical energetic storm particle events observed at shock waves driven by CMEs, especially when the injection energy is high like is typically the case for quasi-perpendicular shocks \cite{tylka05,zank06}. A similar conclusion may hold for the shock waves bounding the CIR.  
We note that in our simulations we avoid any injection problem by considering protons of sufficiently high energy only. 
We leave it for future work to study how our results change when e.g., continuously injecting the low-energy particles from the supra-thermal tail of the solar wind.

The double-enhancement intensity structure never forms at 1 AU in our case~2 and~3 simulations, where we use a cross-field diffusion magnitude comparable to that of Dwyer et al.~\cite{dwyer97}. This may be due to strong turbulence variations near and within the CIR that are not captured by our cross-field diffusion model. 
More specifically, the plasma near the reverse shock is often characterised by significant magnetic fluctuations \cite{tsurutani95, horbury99}, while inside the CIR, the turbulent fluctuation of the magnetic field typically decreases towards the SI \cite{tsurutani82,dwyer97}.  
More recently, Strauss et al.~\cite{strauss16} showed the decrease in turbulence at the SI by computing the variance of the in-situ measured magnetic field components for a CIR. 
In addition, these authors argue that a tangential magnetic discontinuity like the SI may produce an anisotropic cross-field diffusion tensor, by strongly damping magnetic field fluctuations in the direction perpendicular to the discontinuity. 
As a consequence, the SI would act as an efficient diffusion barrier, while still allowing for strong cross-field diffusion near the CIR boundaries. 
Hence, the model proposed by Strauss et al.~\cite{strauss16} seems to explain both the slow increase of the intensity dip in the CIR shown in \cite{tsurutani82} and the large values of $\lambda_\perp/\lambda_\parallel$ reported in \cite{dwyer97} near the reverse shock. 
In our simulations, a strong cross-field diffusion near the shocks leads to less particle acceleration, owing to the efficiency of the cross-field diffusion at spreading particles away from the shock wave. 

As mentioned above, in addition to different shock properties, this apparent inconsistency between our simulation results and in-situ observations may be due to the fact that we are considering 4 MeV protons, whereas Dwyer et al.~\cite{dwyer97} obtain their results for helium with lower energies per nucleon. 
The ion parallel mean free path typically decreases with decreasing rigidity \cite{droge00,droge03}, and therefore low-energy particles would be more efficiently trapped near the shock waves. 
In addition, the strong levels of turbulence observed near CIR shocks might reduce the parallel mean free path of the particles significantly, hence both increasing $\lambda_\perp/\lambda_\parallel$ and facilitating  multiple shock encounters.
Moreover, we note that cross-field diffusion might have a non-negligible pitch-angle dependence, as discussed in \cite{strauss15}, which could also influence the scattering of the particles near the shock waves.
Further investigation of these issues is required, as it will eventually lead to a better understanding of particle transport in the complex turbulent environments that characterise large scale interplanetary structures like CIRs. The strongly changing turbulence properties found in the different parts of a CIR, and the possibility to have in-situ measurements of these regions, make CIRs the ideal laboratories to study the transport of particles perpendicular to the mean magnetic field.

\section*{Acknowledgments}
N. Wijsen is supported by a PhD Fellowship of the Research Foundation Flanders (FWO). The computational resources and services used in this work were provided by the VSC (Flemish Supercomputer Center), funded by the Research Foundation Flanders (FWO) and the Flemish Government, department EWI. The work at KU Leuven was done in the framework of the projects GOA/2015-014 (KU Leuven), G.0A23.16N (FWO-Vlaanderen) and C 90347 (Prodex). Funding for this work at University of Barcelona was partially provided by the Spanish MINECO under project MDM-2014-0369 of ICCUB (Unidad de Excelencia `Mar\'ia de Maeztu'). The work at University of Helsinki was carried out in the Finnish Centre of Excellence in Research of Sustainable Space (Academy of Finland grant numbers 312390 and 312351). 

\section*{References}
\providecommand{\newblock}{}

\end{document}